\documentstyle[aps,prl,twocolumn]{revtex}
\begin{document}                % INITIALIZE - DONT CHANGE
\draft
\title{Effective lowering of the dimensionality in strongly
correlated two dimensional electron gas}
\author{L. B. Ioffe\cite{landau}, D. Lidsky, B. L. Altshuler\cite{MIT} }
\address{Department of Physics, Rutgers University, Piscataway, NJ 08855}
\maketitle
\begin{abstract}
We study a system of fermions interacting with a gauge field which can be
used to describe either spin liquid or $\nu=1/2$ Quantum Hall state.
We propose a generalized model with a dimensionless parameter $N$.
We evaluate the properties of the model in both limits $N \gg 1$ and $N \ll 1$
and deduce the properties of the model in the most physically intersting case
of $N=1,2$.
At $N \ll 1$ the motion of the fermions becomes one dimensional.
This allows us to obtain the fermion Green function and response functions
applying bozonization method in this limit.
\end{abstract}

\narrowtext

Recently the problem of two dimensional (2D) fermions interacting with a
gauge field has become a subject of extensive research.
The low energy behavior of this model is often believed to describe the
properties of two strongly interacting electron systems:
the normal state of the high
$T_c$ cuprates \cite{bza,ioffelar,lee} and the $\nu=1/2$ state in
the Quantum Hall (QH) Effect \cite{hlr,kalmeyer}.

The localized spins in high $T_c$ are assumed to form a gapless
spin liquid \cite{anderson} which does not break any symmetry
( the so-called RVB state ).
It is convenient to describe the excitations of this state in terms of
chargeless fermions with spin 1/2 ( spinons ): ${\bf S}=f^{\dagger}_\beta
\sigma_{ \beta \alpha} f_\alpha$, where $f_\alpha$ is fermion destruction
operator \cite{bza}.
To ensure that each site is occupied by one and
only one spin, it is necessary to couple these fermions to the
gauge field.
Formally, this gauge field appears after a Hubbard-Stratanovich decoupling
of the spin-spin interaction \cite{bza,ioffelar}:
$H=\sum {\bf S}_i {\bf S}_j$.

The 2D electron gas in the QH regime acquires
very special properties in the vicinity of the filling factor $\nu=1/2$.
Attaching two flux quanta to each electron, one maps this problem onto
the problem of fermions interacting with a fluctuating 'magnetic' field of
zero average at $\nu=1/2$ \cite{hlr,kalmeyer}.
Neglecting these fluctuations, one ends up with a gapless Fermi liquid.
The experimentally observed non-zero conductivity \cite{exp}
around $\nu=1/2$ means that this Fermi liquid is a reasonable
starting point.
To complete the theory one must take into account the fluctuations
of the gauge field.

Both problems can be reduced to the Hamiltonian:
\begin{equation}
H= - \frac{1}{2m}f_\sigma^\dagger (\nabla -i{\bf a})^2 f_\sigma +
\frac{\chi}{2}
(\nabla \times {\bf a})^2
\label{model}
\end{equation}
where ${\bf a}$ is a transverse vector potential ($\nabla {\bf a}=0$),
$\sigma=1..N$.
For the spin polarized QH state $N=1$, whereas for the RVB state $N=2$,
corresponding to two possible spin polarizations.
The Hamiltonian (\ref{model}) correctly describes only low energy modes:
fermions in the vicinity of the Fermi line and transverse gauge field.
The high energy modes ( fermions deep in the Fermi sea, longitudinal
gauge field ) were already integrated out in (\ref{model}) resulting in
the stiffness of the gauge field (last term in (\ref{model})) and the
fermion mass renormalization.
The interaction mediated by the longitudinal gauge field is screened and
becomes short ranged, so it can be omitted from the Hamiltonian (\ref{model})).
In case of the QH effect, there is also a coupling between  transverse and
longitudinal gauge fields which gives an additional contribution to the
stiffness $\chi$.

The main difficulty associated with (\ref{model})  is the
singularity of the interaction mediated by the gauge field at low energy
$\omega$ and momentum $k$ transfer.
Fermions near the Fermi line result in a Landau damping \cite{llandau}
($\propto \frac{|\omega|}{k}$), leading to an overdamped dynamics of the gauge
field described by the correlator \cite{ioffelar}
$\langle a_\mu a_\nu \rangle_{k,\omega}$ which we write in Matsubara
(Euclidean) formalism:
\begin{equation}
\langle a_\mu a_\nu \rangle_{k,\omega} =
(\delta_{\mu\nu}-\frac{k_\mu k_\nu}{k^2} ) D(\omega,k)
	= \frac{\delta_{\mu\nu}-\frac{k_\mu k_\nu}{k^2} }
	{\frac{Np_0|\omega|}{2\pi |k|} + \chi k^2}
\label{D}
\end{equation}
Here $p_0$ is inverse curvature of the Fermi line.
For a non-circular Fermi line $p_0$ varies along the Fermi line.
In this case  one has to evaluate curvature at a point where the
tangent to the Fermi line  is parallel to ${\bf k}$.

The model (\ref{model}) has been discussed repeatedly in the
framework of $1/N$ expansion
\cite{ioffelar,lee,ioffekivelson,altshulerioffe,polchiski,wilczek}.
At finite temperatures the static part of gauge field fluctuations
($\omega=0$ in (\ref{D})) dominates \cite{altshulerioffe}, i.e. the
contributions from $\omega \neq 0$ fields are small in $1/N$.
The Green's function of the fermions evaluated in a static approximation
shows that coherence effects are suppressed by magnetic field fluctuations;
as a result, the particle moves semiclassically: for large times only a
small region around the classical trajectory determines the Green's function.

The relevance of $1/N$ expansion for the interesting cases of $N=1,2$ was
questioned in \cite{wilczek}.
Here we shall consider the behavior of the model in both limits $N \rightarrow
\infty$ and $N \rightarrow 0$ and discuss the implications for
$N=1,2$ \cite{khveshenko}.
The limit $N \rightarrow 0$ is realized in the generalized model with
$M$ species of photons.
Such model may describe a spin liquid with $M$ sublattices.
Since in this model the effective correlator (\ref{D}) is multiplied by $M$,
it can be  also described by Eqs. (\ref{model},\ref{D}) with modified
parameters: $\tilde{N}=N/M$, $\tilde{\chi}=\chi/M$.
In the following we shall drop the tildes and study the model for general
values of $N$ and $\chi$.
As will be clear below, the value of the stiffness $\chi$ sets the relevant
length scale $l_0$
\begin{equation}
l_0^{-1} = \left(\frac{1}{3\sqrt{3}} \right)^3
	\frac{2 v_F^2}{\pi^2 p_0 \chi^2N},
\label{l_0}
\end{equation}
at shorter scales the effects of gauge interaction are
not important, whereas the qualitative behavior at larger scales
is governed by the value of $N$.

The static approximation is valid at any temperature $T$ provided that
$N \gg 1$.
In the opposite limit the fermion motion becomes qualitatively
different: it is strongly non-classical and even more one-dimensional.
At $N \ll 1$ the fermion Green's function is given by a $1\!+\!1$ dimensional
integral:
\begin{equation}
G({\bf p},\epsilon)=\int G(x,t) e^{i\epsilon t - i(|{\bf p}| - p_F)x} dt\, dx,
\label{G}
\end{equation}
\[
G(x,t)=\frac{i}{2\pi (x-i v_Ft)} \exp \left(\frac{-\Gamma(2/3) l_0^{-1/3}|x|}
	{(|x|-isgn(x)v_Ft)^{2/3}} \right),
\]
Clearly, the fermion density of states
$\rho = Im \int G(0,t) \times$ $e^{i\epsilon t} dt$ remains the same as
for non-interacting fermions.
At the same time, the probability to move along the classical path
$x=v_Ft$ decays exponentially;
the particle is smeared over a distance $r \propto t^{2/3}$
and the velocity of the wave packet vanishes at $t \rightarrow
\infty$.
Therefore, due to gauge field fluctuations, a single fermion can
not propagate.

However, a particle-hole pair ( which is neutral with respect to the
gauge field ) propagates if the momentum of the relative motion
is small: $k \ll k_\omega=\left( \frac{ N \omega p_0}{\chi} \right)^{1/3}$.
In this case the absorption ( the imaginary
part of the fermion density correlator ) at $k \ll k_\omega$ remains the same
as in the non-interacting Fermi gas: $Im R(\omega,k) \propto \omega/|k|$.

The particles with large relative velocities become
essentially independent and propagate poorly.
This effect suppresses the absorption at large wave vectors $k \gg k_\omega$,
we estimate it as $Im R(\omega,q) \leq \omega(p_F l_0)$.

Increasing $N$ results in a crossover to the state where
static fluctuations of the gauge field dominate at finite temperatures.
At $T=0$ the most important effect of the $1/N$ corrections is the anomalous
power law behavior of the density ( and spin ) correlators at $k=2p_F$
and the backward scattering amplitudes, these exponents are proportional to
$1/N$.
The power laws cross over to an exponential behavior at $N \rightarrow 0$.

Now we sketch the derivation of these results.
To the first order in the interaction with the gauge field the Green's
function is
$
G^{(1)-1}(\epsilon,{\bf p}) = i\epsilon-v_F(|p|-p_F)-\Sigma^{(1)}(\epsilon),
$
\begin{equation}
\begin{array}{ll}
\Sigma^{(1)}(\epsilon)= v_F^2 \int \frac{D(\omega,k) d\omega d^2k}
	{i(\epsilon + i\omega) - v_F(|p+k|-p_F)} =
-i \left| \frac{\omega_0}{\epsilon} \right|^{1/3} \epsilon
\\
\omega_0=\left(\frac{3}{2}\right)^3 \frac{v_F}{l_0} \propto \frac{1}{N}
\end{array}
\label{Sigma1}
\end{equation}
for the idea of the derivation see \cite{lee,reizer}.

At $N \gg 1$ diagrams with crossings are negligible.
Consider, e.g. the diagram for the self energy shown in Fig. 1a, it gives

\begin{eqnarray}
\Sigma^{(2)}({\bf P} &&) = v_F^4
\int G^{(1)}({\bf P+K_1})
    G^{(1)}({\bf P+K_2}) \times
\label{Sigma2e}
\\
&& G^{(1)}({\bf P+K_1+K_2})D(K_1)D(K_2) (d^3K_1d^3K_2)
\nonumber
\end{eqnarray}
where $D$ and $G^{(1)}$ are given by (\ref{D},\ref{Sigma1}) and $\bf P$,
$\bf K$ denote 3D vectors ( e.g. ${\bf P}=(\epsilon, {\bf p})$ ).

To evaluate the self energy (\ref{Sigma2e}) we
introduce the components $k_{\parallel}$ ($k_\perp$) of the momenta
$k$ parallel ( perpendicular ) to $\bf p$.
The product of three Green's function in (\ref{Sigma2e}) decreases rapidly
at $v_F k_\parallel \gg \Sigma(\epsilon)$, so
the main contribution to (\ref{Sigma2e}) comes from the range of small
($k_{\parallel}\lesssim l_0^{-1}\left(\frac{\epsilon}{\omega_0}\right)^{2/3}$)
momenta.
Both $G$ and $D$ decrease only when $k_{\perp} \gtrsim k_\epsilon$, so
the main contribution to (\ref{Sigma2e}) comes from the range
$k_{\perp} \lesssim k_\epsilon \sim
k_\parallel \left(\frac{\omega_0}{\epsilon}\right)^{1/3} \gg k_{\parallel}$.
Neglecting the $k_{\parallel}$-dependence of the gauge field propagator
we get
\begin{eqnarray}
\Sigma^{(2)}(\epsilon) &&= v_F^2 \int \frac{A(\omega_1,\omega_2)
	(d\omega_1d\omega_2dk_{\perp 1}dk_{\perp 2})}
    {A^2(\omega_1,\omega_2)-k_{\perp 1}^2k_{\perp 2}^2/m^2} \times
\nonumber \\
&&D(\omega_1,k_{\perp 1}) D(\omega_2,k_{\perp 2})
\label{Sigma2i}
\\
A(\omega_1,\omega_2) &&= v_F(p_F-p) +
\label{A} \\
i \beta &&\left(
|\epsilon+\omega_1 +\omega_2|^{2/3} + |\epsilon + \omega_1|^{2/3} + |\epsilon +
\omega_2|^{2/3} \right)
\nonumber
\end{eqnarray}
In the limit $N \gg 1$ the integral over $k_{\perp}$ can be performed
with logarithmic accuracy:
\begin{equation}
\Sigma^{(2)} = - i c \beta \left( \frac{\ln N}{4\pi N} \right)^2
	\epsilon \left|\frac{\omega_0}{\epsilon}\right|^{2/3},
\hspace{0.25in} c \approx 2.16
\label{Sigma2}
\end{equation}
Note that the ratio of $\Sigma^{(2)}/\Sigma^{(1)}$ is governed only by $N$,
at $N \gg 1$ $\Sigma^{(2)} \ll \Sigma^{(1)}$.
This smallness can be traced back to the term $k_{\perp 1}^2k_{\perp 2}^2$
in the denominator of (\ref{Sigma2i}) caused by the crossing in this diagram.

Since each crossing results in  an additional factor $1/N^2$,
only the diagrams with the minimal numbers of crossings are important at
$N \gg 1$.
 From the theory of localization such diagrams are known to
include the ladder-like pieces in particle-hole and particle-particle
channels.

The sum of the ladder diagrams for the vertex part shown in Fig 1c
has a singularity at small momenta which is reminiscent of a diffusion
pole in localization problem:
\begin{equation}
\Gamma_D(\omega,\epsilon,k)=
	 \frac{
(\frac{\omega}{2}-\epsilon)\frac{\omega_0}{|\frac{\omega}{2}-\epsilon|^{1/3}}+
	(\frac{\omega}{2}+\epsilon)
		\frac{\omega_0}{|\frac{\omega}{2}+\epsilon|^{1/3}}
	}{
	ivk_{\parallel}sgn(\omega) +
		c'\frac{k_\omega}{p_0} v_F|k_{\perp}|}
\label{Gamma_D}
\end{equation}
where $c'= 0.2262$ and we neglected terms $O(\omega)$ in the denominator.
We insert this vertex correction into the first order diagram for the self
energy and find the contribution  proportional to
$\frac{1}{N} \beta \epsilon^{2/3}$.
Therefore the singularity (\ref{Gamma_D}) has no effect on $\epsilon$
dependence but reduces the power of $1/N$ in the leading term of $1/N$
expansion.
Inserting the renormalized vertex (\ref{Gamma_D}) in the polarization bubble
we again find only small correction ($ \propto |\omega|^{2/3}$) at
characteristic frequencies and momenta of the gauge field
( $k \sim k_\omega$ ).
Thus, at low energies the polarization bubble is not renormalized
significantly proving that the Green function (\ref{D}) gives an asymptotically
exact answer in this limit.
The analogous diagram in the particle-particle ( Cooper ) channel is not
singular.

The vertex parts at the momentum transfer $2p_F$ are more interesting.
Their anomalous energy dependence determines the dependence of spin
and density correlators.
The first order correction is proportional to $\ln\epsilon$
\cite{ioffekivelson}.
The series of logarithms can be summed similarly to the
derivation of doubly logarithmic asymptotics in quantum electrodynamics
\cite{berestetskii}, at small energies and momenta close to $2p_F$ the
vertex has a power law singularity:
\begin{equation}
\Gamma_P(\epsilon,2p_F)  \sim \left( \frac{\epsilon_F}{|\epsilon|}
			\right)^{\frac{1}{2N}} U,
\hspace{0.15in}
\Gamma_P(0,k)  \sim \left( \frac{p_F}{|k|-2p_F} \right)^{\frac{3}{2N}}
\label{Gamma_P}
\end{equation}

Interaction between fermions with momentum transfer close to $2p_F$
can be treated similarly.
Using renormalization group (RG) approach to sum the leading logarithmic
corrections (Fig 1d) to the interaction vertex $U$ we find the RG equation:
\begin{equation}
\frac{dU}{d\xi} = - \left( \frac{3}{4} - \frac{1}{N} \right) U
\end{equation}
Therefore, for sufficiently small bare interaction and $N \gg 1$,
it decreases and becomes negligible at large scales.
In this regime the spin and charge susceptibilities are given by the
polarization bubble with renormalized vertices (\ref{Gamma_P}).

Now we turn to the limit  $N \rightarrow 0$.
We can neglect the dependence of the fermionic Green's function
on the transverse momenta even in  higher order diagrams with crossings,
e.g. we can neglect the term $k_{\perp 1}^2k_{\perp 2}^2$ in the denominator
of (\ref{A}) for the diagram in Fig 1a.
Then $k_{\perp}$ enters only via $D(\omega,k)$ and integrals over them
factorize.
The resulting series of diagrams coincides, order by order, with a 1D
theory with the action
\begin{equation}
S=\int \left( \bar{\Psi}_{-\omega,-k}^a(i\omega -v_Fk)\Psi_{\omega,k}^a +
	\frac{v_F g |\rho_{\omega,k}|^2 }{|\omega|^{1/3}} \right)
	(dkd\omega)
\label{1DS}
\end{equation}
where $\rho_{t,x}=\bar{\Psi}_{t,x}^a\Psi_{t,x}^a$ is the density operator and
$g=\frac{2}{3} \left(2\pi \omega_0\right)^{1/3}$ is
effective interaction constant of this one dimensional theory.

The diagrams with Fermi loops ( e.g. in Fig. 1b ) are already taken into
account in the gauge field propagator (\ref{D}).
As we show below, the higher order insertions in the loop of the Fermi field
does not change the propagator (\ref{D}) at typical $\omega$ and $k$.
However, a perturbative treatment of model (\ref{1DS}) generates
the loops.
To get rid of them we use the replica trick - take the limit of zero number
of components of the Fermi field: $a=1..s$, $s\rightarrow 0$.

Finally, the retarded interaction between the fermions provides a natural cut
off of the ultraviolet divergencies:
regularization of the 1D theory in the time direction is
more natural here than the conventional regularization
in the space direction.

The 1D model (\ref{1DS}) can be solved by bosonization
\cite{fradkin}.
As usual, we introduce the bose field $\phi$ with the action quadratic in bose
fields
\begin{equation}
S_B=\frac{1}{2} \int \left( (\omega^2+k^2)(\phi_a)^2 +
\frac{g}{\pi} |\omega|^{5/3} \phi_a \phi_b \right) (d\omega dk)
\label{S_B}
\end{equation}
This allows us to evaluate the Green's function:
\begin{equation}
F_{ab}(\omega,k) = \langle \phi_a \phi_b \rangle =
	\frac{1}{\omega^2+k^2} \delta_{ab} -
	\frac{g |\omega|^{5/3}}{\pi (\omega^2 + k^2)^2 }
\label{F}
\end{equation}
The original fermions are related to the bose field by
\begin{equation}
\Psi=\frac{1}{\sqrt{2\pi a}} \exp\sqrt{\pi}(\phi^{*}+i\phi)
\label{Psi}
\end{equation}
where $\phi^{*}$ is the dual field: $\partial_t \phi^{*}=\partial_x \phi$.
All correlators of the original Fermi fields are reduced to the gaussian
integrals over the bose fields.
Evaluating  these integrals we get the fermion Green's function (\ref{G}).

At small momenta we note that at $k\leq k_\omega$ the main
contribution to the absorption comes from fermion hole pairs close to the
Fermi surface with $v_F k_{\parallel} \sim \Sigma(\omega)$ and perpendicular
$k_{\perp} \approx k \ll k_\omega$.
The momentum perpendicular to $\bf p$ can be
ignored and the interaction between fermion and hole can be described within
the 1D theory.
Effects of the interaction cancel exactly in the density correlator
$ \langle \rho \rho \rangle = \langle (\bar{\Psi}_a \Psi_b)
(\bar{\Psi}_b \Psi_a) \rangle $:
the absorption at these wave vectors is exactly the same as for
non-interacting fermions.
This also proves that the interaction does not renormalize
$D(\omega,k)$ in the important range of $\omega$ and $k$.
Thus, we have shown that the gauge field propagator (\ref{D}) is not
renormalized in both limits $N \gg 1$ and $N \ll 1$, so we conjecture that it
is not renormalized for any $N$.

As we have seen, at $N \gg 1$, scattering with momentum transfer $2p_F$ is
enhanced.
This effect is even stronger at $N \ll 1$.
This scattering is known to be renormalized to infinity in a case of a
Lutinger liquid with repulsion \cite{fisherk}.
In a 1D model (\ref{1DS}) the enhancement is much stronger than in a
canonical Lutinger liquid discussed in \cite{fisherk}; namely,
the bosonization approach gives the backscattering probability
\begin{equation}
\tau_B^{-1} \sim \tau^{(0)-1} \exp \left( \frac{3}{2\pi g
	|\omega|^{1/3}} \right)
\label{tau}
\end{equation}
where $1/\tau^{(0)}$ is the backscattering probability of free electrons.
The difference from the Luttinger liquid originates from the power
singularity in (\ref{1DS}).
The exponentially large amplitude of the backscattering is due to a
cancellation of the diagrams in the 1D theory which follows from a generalized
Ward identity \cite{dzyaloshinskii}.
This cancellation is exact  only at $N \rightarrow 0$, so we expect that
the result (\ref{tau}) crosses over to a power law behavior (\ref{Gamma_P})
at finite $N$ with an exponent that tends to infinity when $N \rightarrow 0$.

We believe that the effect of backscattering enhancement was observed
experimentally in \cite{stormer} where it was found that for
$\nu = 1/2$ the transport relaxation rate is about two orders of
magnitude larger than in the state without magnetic field while the total
relaxation rate ( determined from Shubnikov-De Haas oscillations )
remained of the same order.
With no magnetic field the total scattering rate, was much larger than the
transport rate implying a very small probability of backscattering.
This probability can be dramatically enhanced by the mechanism we have
proposed in this paper.

When the magnetic field varies away from $\nu=1/2$, the fermions see an
incremental uniform field; its influence on the dimension reduction is still
an open question.

In conclusion, we have shown that two dimensional fermions interacting with
a gauge field may exhibit one dimensional behavior.
This conclusion seems consistent with a qualitative picture proposed by
Anderson \cite{anderson2}.
This approach may lead to the quantitative description of both the RVB and
$\nu=1/2$ states.

We are grateful to B. Halperin, P. A. Lee, A. Millis and A. Zamolodchikov
for useful discussions.

\begin{figure}
\caption{Typical high order diagrams. a. Simplest high order diagram for
the self energy with crossings.
b. Loop insertion in the gauge field line that should be excluded from 1D
theory.
c. At $N \ll 1$ only ladder diagrams are important for vertex
renormalization.
$k \approx 0$ corresponds to vertex $\Gamma_D$, $k \approx 2p_F$ corresponds to
vertex $\Gamma_P$.
Filled triangle denotes renormalized vertices $\Gamma_D$, $\Gamma_P$.
d. Renormalization of the interaction ( denoted by square )
with momentum transfer about $2p_F$.
}
\end{figure}


\begin{references}
\bibitem[*]{landau}also at Landau Institute of Theoretical Physics, Moscow,
Russia.
\bibitem[\dagger]{MIT} Department of Physics, MIT, Cambridge, MA 02139.
\bibitem{bza} G. Baskaran at al, Sol. St. Comm. {\bf 63}, 973 (1987);
A. E. Ruckenstein at al, Phys. Rev. {\bf 36}, 857 (1987);P. B. Wiegmann,
Phys. Rev. Lett. {\bf 60}, 821 (1988); X. G. Wen at al, Phys. Rev. {\bf B 39},
11413 (1989).
\bibitem{ioffelar} L. B. Ioffe and A. I. Larkin, Phys. Rev. {\bf B 39},
8988 (1989).
\bibitem{lee} P. A. Lee, Phys. Rev. Lett. {\bf 63}, 680 (1989);
N. Nagaosa and P. A. Lee, Phys. Rev. {\bf B 46}, 5621 (1992).
\bibitem{hlr} B. Halperin at al, Phys. Rev. {\bf B 47}, 7312 (1993).
\bibitem{kalmeyer} V. Kalmeyer and S. C. Zhang, Phys. Rev. {\bf B 46}, 9889
(1992).
\bibitem{anderson} P. W. Anderson, Science, {\bf 235}, 1196 (1987).
\bibitem{exp} H. W. Jiang at al, Phys. Rev. {\bf B 40}, 12013 (1989).
\bibitem{llandau} L. D. Landau, Sov. Phys. JETP {\bf 10}, 25 (1946).
\bibitem{ioffekivelson} L. B. Ioffe, S. A. Kivelson, A. I. Larkin,
Phys. Rev. {\bf B44} 12537-12543 (1991).
\bibitem{altshulerioffe} B. L. Altshuler, L. B. Ioffe, Phys Rev Lett, {\bf 69},
2979 (1992).
 \bibitem{polchiski} J. Polchinski, unpublished (1993).
\bibitem{wilczek} C. Nayak, F. Wilczek, unpublished (1993).
\bibitem{khveshenko} It is likely that the idea of eikonal approximation
proposed recently by D. V. Khveshchenko and P. C. E. Stamp in Phys. Rev. Lett.
{\bf 71}, 2118 (1993) can be justified in $N \rightarrow 0$ limit.
\bibitem{reizer}M. Reizer, Phys. Rev. {\bf 40}, 11571 (1989).
\bibitem{berestetskii} V. B. Berestetskii, E. M. Lifshitz, L. P. Pitaevskii,
``Quantum Electrodynamics'', \S 136, Pergamon Press (1980).
\bibitem{fradkin} See, e.g., E. Fradkin,
``Field Theories of Condensed Matter Systems'', Addison-Wesley (1991).
\bibitem{fisherk}C. Kane and M. P. A. Fisher, Phys. Rev. Lett. {\bf 68}, 1220
(1992); A. I. Larkin and P. A. Lee, Phys. Rev. {\bf B 17}, 1596 (1978).
\bibitem{dzyaloshinskii} I. E. Dzyaloshinkii, A. I. Larkin, Sov. Phys. JETP,
{\bf 38}, 202 (1974).
\bibitem{stormer} R. R. Du, at al, Bell Labs preprint (1993)
\bibitem{anderson2} P. W. Anderson, Phys. Rev. Lett. {\bf 64}, 1839 (1990);
{\bf 65}, 2306 (1990); {\bf 66}, 3226 (1991); {\bf 67}, 2092 (1991).
\end{references}
\end{document}